\documentclass[11pt]{article}
\usepackage[latin1]{inputenc}
\usepackage[english]{babel}
\usepackage[namelimits]{amsmath}
\usepackage{amssymb}
\usepackage{amsmath}
\usepackage{amsthm}
\newcommand{\ac}{\'}

\begin{document}
\title{An inhomogeneous universe with thick shells and without a cosmological constant}
\author{Stefano Viaggiu
\\
Dipartimento di Matematica, Universit\'a "Tor Vergata"',\\ 
Via della Ricerca Scientifica, 1\\
Rome, Italy 00133,\\
viaggiu@axp.mat.uniroma2.it}
\date{\today}\maketitle

\begin{abstract}
We build an exact inhomogeneous universe composed of
a central flat Friedmann zone up to a small redshift $z_1$,
a thick shell made of anisotropic matter, 
an hyperbolic Friedmann metric up to the scale where dimming galaxies are observed
($z\simeq 1.7$) that can be matched to a
hyperbolic Lema\^{i}tre-Tolman-Bondi spacetime to best fit
the WMAP data at early epochs. 
We construct a general framework which permits us to consider
a non-uniform clock rate for the universe. As a result, both for a uniform 
time and a uniform Hubble flow, the deceleration parameter extrapolated by the
central observer is always positive. Nevertheless, by taking a non-uniform
Hubble flow, it is possible to obtain a negative central deceleration parameter,
that, with certain parameter choices, can be made the one observed currently.
Finally, it is conjectured a possible physical mechanism to justify a non-uniform
time flow.
\end{abstract}
PACS numbers: 98.80.-k,98.80.Jk,95.36.+x,04.20.-q

\section{Introduction}
Supernovae type Ia (SNIa) observations of the past decade seem to indicate 
an accelerating universe (\cite{1,q,2}). In the standard approach with 
the Friedmann-Lema\^{i}tre models (FRLW),
an accelerating universe invokes the 
presence of a large amount of the so called dark energy. In the 
FRLW picture, this dark energy is given by the cosmological
constant. The dark energy represents a puzzle and perhaps the biggest problem
in modern cosmology. In fact, 
a direct detection of a cosmological constant is
still lacking. In the last decade, many attempts have been made 
(see \cite{3}-\cite{20BB} and references therein) to obtain physically 
sensible inhomogeneous models. 
Some authors
(see for example \cite{13,123,13BB,15,16,20,20BB}) showed that inhomogeneities can generate 
an accelerating universe by using Lema\^{i}tre-Tolman-Bondi (LTB)
metrics (see \cite{m1,21,22}), but several conditions 
must be imposed (see \cite{3}) in order to build regular physically viable models.
In particular, in \cite{15,16} it is shown that LTB metrics can 
mimic the distance-redshift
relation of the FRLW models at least at the third order in a series 
expansion with respect
to the redshift near the center where the observer
is located. More generally, in the LTB
solutions, apparent acceleration in the redshift-distance relation seen by a central
observer can be shown to coexist with a volume average deceleration on a spacelike
hypersurface (see \cite{222}). The assumption of spherical symmetry is (obviously) not
in agreement with
the Copernican principle. In any case, a spherical symmetry can be 
justified as the outcome of a smoothing out with respect to the angles: the metric so
obtained becomes spherical.\\
An accelerating universe
can also be built by averaging inhomogeneities (see \cite{3,3B,3BB,4,5,500,6,7,8,10,11,12})
by means of the techniques depicted in \cite{5,6,7}. The approaches dealing by averaging
on spatial domains are different with respect to the ones dealing with exact spherical
solutions. First of all, averaged cosmologies retain the Copernican principle, although in a
generalized statistical sense. Further, the exact spherically symmetric models try to describe
apparent acceleration by means of 
a large amount of inhomogeneities. Conversely, the
"Copernican" cosmologist introduce inhomogeneities without special symmetries and then
try to understand the modifications to the average evolution from backreaction.
For a review of 
inhomogeneous cosmological models see \cite{d1,d2}.
In particular, we study the idea
developed in Wiltshire's papers \cite{10,11,12}. There, the dimming of the distant 
galaxies is interpreted as a "mirage" effect by means
of a "Copernican" statistical model. This effect is due to the different rate
of clocks located in averaged not expanding galaxies, where the metric is spatially flat,
with respect to clocks in voids where the spatial curvature is negative.
With a negative spatial curvature can be associated a positive quasilocal
energy. This gravitational energy is non-local, according to 
the strong equivalence principle. In this picture the universe is composed by 
a cosmic web of regions evolving asymptotically like an Einstein-de Sitter universe
(our local universe) "matched" with local voids evolving like a Milne universe.
The matching conditions are imposed with a uniform Hubble flow by equating the
radial null sections of the cosmic web. Although this matching is reasonable,
it cannot be enough to define a geometry. In this context, inspired by Wiltshire's idea,
we introduce clock effects, but with an exact spherically symmetric model and without
introducing backreaction. In this way we can build a geometry with a non-uniform
time flow by imposing the usual matching conditions required by general relativity, 
which are missing in 
\cite{10,11,12}. To mimic Wiltshire's idea,
our model is built with a central zone up to a small redshift
$z_1$ ($<100/h\;Mpc$) 
made of a flat Friedmann metric, where the observer is located.
In practice, the inhomogeneities can be taken into
account by glueing together different homogeneous portions of universe (see for example
\cite{20BB}). It should be noticed that cosmological models with only two metrics can
be exhaustively found in \cite{ad}.
By means of a thick shell of anisotropic matter, the flat zone can be smoothly matched
with an  hyperbolic Friedmann metric
up to the zone where dimming galaxies are observed \cite{25}. Finally, we stress that
the hyperbolic metric can be smoothly matched to a hyperbolic
LTB solution
on a comoving boundary surface, according to
WMAP data \cite{26}. Actually, we could choose
the flat Friedmann one (without cosmological constant!), but WMAP predicts a flat metric
only at early times, while at later epochs a negatively curved metric is more appropriate.
Our solution is exact and
no approximations have been made, thus providing a general framework to
analyze the effects of a possible non-uniform time flow. In this context, we show
that, after a formal Taylor expansion of the distance-redshift relation near
the center at $z=0$, a negative deceleration parameter is only compatible 
with a non-uniform Hubble flow within a non-uniform time flow. The physical
content of the thick shell is studied together with the energy conditions.
Finally, it is conjectured that a non-vanishing heat flow term in the
energy-momentum tensor of the thick shell can explain the possibly non-uniform
time flow.\\
In section 2 we introduce the metrics of our model. In section 3
the junction conditions are discussed. In section 4 the matter content
of the thick shell is studied. In section 5 the case of a uniform
time flow is analyzed, while in section 6 
the case of a uniform Hubble flow is presented. Section 7
is devoted to the study of the general case. Finally, section 8
collects some final remarks and conclusions.

\section{The model}
All the astrophysical observations agree with the assumption that 
our "near" local universe is rather inhomogeneous. Nevertheless,
at sufficiently large scales ($\sim 100/h\;Mpc$) the inhomogeneities can be averaged to obtain
a statistically homogeneous universe.
In any case, from the perturbations of the primordial inflation,
a model with an overdense density surrounded by voids seems to be the
most probable. Hence, we can build an inhomogeneous universe 
by glueing together different homogeneous volumes. As a 
consequence, the central region, with the observer
located at the center, is provided by a flat Friedmann metric
\begin{eqnarray}
& &ds_{B}^{2}=-dt_{B}^{2}+a_{B}^{2}(t_B)(d{\eta}_{B}^{2}+{\eta}_{B}^{2}\;
d{\Omega}^2),\nonumber\\
& &a_B(t_B)=a_{Bi}{\left(\frac{t_B}{t_{Bi}}\right)}^{\frac{2}{3}},
\label{1}
\end{eqnarray}
where in (\ref{1}) we have assumed a dust model and $a_{Bi}, t_{Bi}$
are initial values.
With (\ref{1}),
the Hubble flow $H_B$ is given by
\begin{equation}
H_B=\frac{1}{a_B}\frac{da_B}{dt_B}=\frac{2}{3t_B}.
\label{2}
\end{equation}
The zone where the dimming galaxies are observed ($ z\leq 1.7$) is 
modeled with an hyperbolic
Friedmann metric with negative spatial curvature and a common
centre with (\ref{1}).
In appropriate coordinates, the metric can be put in the form
\begin{eqnarray}
& &ds_{F}^{2}=-dt^2+{\overline{a}(t)}^2[d{\eta}_{F}^{2}+
{\sinh}^2{\eta}_{F}\;d{\Omega}^2],\label{3}\\
& &{\overline{H}}_i t=
\frac{{\overline{\Omega}}_i}{2{(1-{\overline{\Omega}}_i)}^{\frac{3}{2}}}
\left(\sinh\xi-\xi\right),\nonumber\\
& &{\overline{a}}(t)=\frac{{\overline{a}}_i{\overline{\Omega}}_i}
{2(1-{\overline{\Omega}}_i)}\left(\cosh\xi-1\right),\nonumber\\ 
& &{\overline{a}(t)}^2{\overline{H}}^2(1-{\overline{\Omega}})=1,\nonumber
\end{eqnarray}
where ${\overline{\Omega}}_i$ is an initial density parameter,
${\overline{H}}_i$ an initial Hubble constant and 
${\overline{a}}_i$ an initial expansion factor to be specified.
An observer in the portion of universe given by (\ref{3}) measures
the observables by means of the comoving time $t$. 
With respect to  this time, an observer in (\ref{3}) measures 
an Hubble flow with a time dependent Hubble constant
$\overline{H}$ given by
\begin{equation}
\overline{H}=\frac{1}{\overline{a}}\frac{d\overline{a}}{dt}=
\frac{2{\overline{H}}_i\sinh\xi}{{\overline{\Omega}}_i{(\cosh\xi-1)}^2}
{\left(1-{\overline{\Omega}}_i\right)}^{\frac{3}{2}}.\label{4}
\end{equation}
The only way to smoothly  match the metrics (\ref{1}) and 
(\ref{3}) is by means of a thick shell living in the region
$z\in[z_1,z_2]$.
Without loss of generality, we can choose the metric of the
thick shell mimic the expression of a
LTB metric
\begin{equation}
ds_{thick}^{2}=-e^{G(\tau,\eta)}{d\tau}^2+
\frac{R_{,\eta}^2(\tau,\eta)}{f^2(\eta)}d{\eta}^2+R^2(\tau,\eta)
d{\Omega}^2,\label{5}
\end{equation}
where $e^{G(\tau,\eta)}$ denotes the lapse function. 
Furthermore, to achieve agreement with WMAP data \cite{26}, 
the universe beyond the dimming zone could be modeled with
an hyperbolic LTB spacetime (see \cite{m1,21,22}) 
\begin{equation}
ds^{2}=-{d{\tilde{\tau}}}^2+
\frac{{\tilde{R}}_{,\tilde{\eta}}^2(\tilde{\tau},\tilde{\eta})}{{\tilde{f}}^2(\tilde{\eta})}
d{{\tilde{\eta}}}^2+{\tilde{R}}^2(\tilde{\tau},\tilde{\eta})
d{\Omega}^2,\;\;{\tilde{f}}^2(\tilde{\eta})>1.\label{6}
\end{equation}
In fact, WMAP forces us to conclude that at early epochs the LTB metric must 
approach a flat one and, as a result, we must impose the condition that
the density parameter ${\Omega}_m$  approaches unity at early times
(recombination era), i.e. ${\Omega}_m({\tilde{\tau}}_{rec})\rightarrow 1$. According to \cite{31bis},
at later epochs the metric could as well be taken to have a negative
spatial curvature.

\section{Matching conditions}
First of all, it should be noticed that the matching 
of the Friedmann metrics of this paper can be obtained only by taking a thick shell. 
We perform the matching along
comoving surfaces (the boundary of the thick shell) given by
\begin{equation}
{\eta}_{B}={\eta}_{B(1)},\;\eta={\eta}_{(1)},\;
\eta={\eta}_{(2)},\;{\eta}_F={\eta}_{F(2)},\label{7}
\end{equation}
where the subscripts $(1)-(2)$ denote the boundaries of the shell.
For the thick
shell, the continuity of the first and the second fundamental form
\cite{f1,f2} leads to
\begin{eqnarray}
& &{\left(\frac{dt_B}{d\tau}\right)}_{(1)}=
e^{\frac{G(\tau,{\eta}_{(1)})}{2}},
{\left(\frac{dt}{d\tau}\right)}_{(2)}=e^{\frac{G(\tau,{\eta}_{(2)})}{2}},
\label{8}\\
& &R(\tau,{\eta}_{(1)})={\eta}_{B(1)}a_B(t_B),\nonumber\\
& &R(\tau,{\eta}_{(2)})={\overline{a}}(t)\sinh{\eta}_{F(2)},\label{9}\\
& &f({\eta}_{(1)})=1,\;f({\eta}_{(2)})=\cosh{\eta}_{F(2)},\label{10}\\
& &G_{,\eta}(\tau,{\eta}_{(1)})=G_{,\eta}(\tau,{\eta}_{(2)})=0.\label{11}
\end{eqnarray}
Together with equations (\ref{8})-(\ref{11}), we have the "gauge" condition
\begin{equation}
\frac{dt}{dt_B}=J(\xi),\;\;\overline{H}(\xi) = \alpha(\xi) H_{B}(t_B).
\label{12}
\end{equation}
The function $J(\alpha(\xi))$ depends upon the chosen function $\alpha$.
With condition (\ref{11}), the heat flow vanishes at the boundaries
of the thick shell (see the next section). To integrate the system 
(\ref{8})-(\ref{11}), we can fix an expression for $G$ satisfying 
equation (\ref{11}) and the relations $t=t(\tau,\eta), t_B=t_B(\tau,\eta)$
that satisfy equation (\ref{8}). In this way, thanks to the equations
(\ref{9}), the behaviour of $R(\tau,\eta)$ is fixed at the boundaries
$(1)-(2)$, and, as a result, we have the freedom to choose $R(\tau,\eta)$
inside the shell and so also the function $f$  with conditions (\ref{10}).\\
For a smooth matching between (\ref{3}) and (\ref{6}) on a comoving boundary surface
${\eta}_F={\eta}_{F(3)},\;{\tilde{\eta}}={\tilde{\eta}}_{(3)}$ we have
\begin{eqnarray}
& &\tilde{\tau}=t,\label{2222}\\
& &\tilde{R}(\tilde{\tau},{\tilde{\eta}}_{(3)})={\overline{a}}(t)\sinh{\eta}_{F(3)},\label{13}\\
& &\tilde{f}({\tilde{\eta}}_{(3)})=\cosh{\eta}_{F(3)},\label{15}
\end{eqnarray}
It should be noticed that there exists a relationship between the metrics (\ref{5}) and (\ref{6}).
First of all, at the recombination era (early epochs) we have:
\begin{equation}
{\tau}_{rec}\simeq{\tilde{\tau}}_{rec},\;G({\tau}_{rec},\eta)\simeq 0,\;
R({\tau}_{rec},\eta)\simeq{\tilde{R}}({\tilde{\tau}}_{rec},{\tilde{\eta}}).
\label{wer}
\end{equation}
Furthermore, thanks to the matching conditions (\ref{8}) and 
(\ref{2222}), we have (remember that ${\eta}_{(2)}$ is a constant) 
\begin{equation}
e^{G(\tau,{\eta}_{(2)})}d\tau^2=g(\tau) d\tau^2 = d \tilde\tau^2,
\end{equation}
Finally, without loss of generality, we could also set $\eta=\tilde\eta$
in (\ref{5}) and (\ref{6}).\\
Since the matching conditions have been discussed, we can give the formal 
expressions for the angular distance $d_A$ and the distance luminosity $d_L$
where $d_L=d_A{(1+z)}^2$
(see \cite{dl1,dl2,dl3}).
In fact, thanks to conditions (\ref{8})-(\ref{11}) and 
(\ref{2222})-(\ref{15}), we obtain (see \cite{20BB})
\begin{eqnarray}
& &d_A=a_B(t_B){\eta}_B,\;\;\;\;\;\;\;\;z\leq z_1,\label{17}\\
& &d_A=R(\tau,\eta),\;\;\;\;\;\;\;\;\;\;\;\;z\in [z_1,z_2],\nonumber\\
& &d_A=\overline{a}(t)\sinh{\eta}_F,\;\;\;\;\;z\in [z_2,z_3],\nonumber\\
& &d_A=\tilde{R}(\tilde{\tau},\tilde{\eta}),\;\;\;\;\;\;\;\;\;\;\;\;z\geq z_3,\nonumber
\end{eqnarray}
where, $z_3$ ($>1.7$) represents the "starting point" of the LTB metric.
For our purposes, we are interested in the patch of universe where the 
dimming galaxies are observed, i.e. the third of equations (\ref{17}).

\section{Energy-momentum tensor for the thick shell}
In this section we study the metric (\ref{5}). The most general energy-momentum
tensor $T_{ab}$ compatible with it is
\begin{eqnarray}
& &T_{ab}=EV_aV_b+P_{\perp}\left[W_aW_b+L_aL_b\right]+\nonumber\\
& &+P_{\eta}S_a S_b+K\left[V_aS_b+S_aV_b\right],\label{18}
\end{eqnarray}
with
\begin{eqnarray}
& &V_a=\left[-e^{\frac{G}{2}},\;0,\;0,\;0\right],\label{19}\\
& &W_a=\left[0,\;0,\;R,\;0\right],\nonumber\\
& &L_a=\left[ 0,\;0,\;0,\; R\sin\theta\right],\nonumber\\
& &S_a=\left[0,\;\frac{R_{,\eta}}{f},\;0,\; 0\right],
\end{eqnarray}
where $E$ is the energy-density, $P_{\eta}$ the radial pressure,
$P_{\perp}$ the tangential pressure, $K$  being the "heat flow term"
or radial energy flux.
In particular, Einstein's equations for $K$ give
\begin{equation}
K= -\frac{f R_{,\tau}G_{,\eta}}{R R_{,\eta}e^{\frac{G}{2}}}.
\label{20}
\end{equation}
The regularity conditions require
that $(R_{,\eta}, R_{,\tau}, R, f)\neq 0$.
Hence, from equation (\ref{20}),
it follows that $K=0$ if and only if
$G_{,\eta}= 0$. Therefore,
a non trivial lapse function is only compatible with a non-vanishing
energy flux. If we take the most general expression for a spherically
symmetric metric that is
\begin{equation}
ds^2=-e^{G(\tau,\eta)}d{\tau}^2+\frac{A^{2}_{,\eta}(\tau,\eta)}{f^2(\eta)}d{\eta}^2+
B^2(\tau,\eta)d{\Omega}^2,
\label{21}
\end{equation}
the energy flux for (\ref{21}) vanishes if and only if
\begin{equation}
G_{,\eta}=2{(\ln B_{,\tau})}_{,\eta}-
2{(\ln A_{,\tau})}_{,\eta}\frac{B_{,\eta}}{B_{,\tau}}.
\label{22}
\end{equation}
To the best of our knowledge, the only non-static metric with a non-barotropic
equation of state satisfying equation (\ref{22}) is the Stephani metric 
\cite{ext}. Remember that a time dependent spherical matter cannot have a barotropic
equation of state ($E=E(P)$) within a finite radius \cite{ase}. It is a simple matter
to see that the spherically symmetric Stephani space-time cannot satisfy the matching
conditions (\ref{8})-(\ref{11}). As a result, a link between the 
heat flow and a non-trivial lapse function can be conjectured, 
at least for spherically symmetric space-times.
Should this argument be correct, we would have a possible physical mechanism to
generate clock effects.\\
In what follows, starting from conditions (\ref{8})-(\ref{11}),
we analyze the possibility to build a physically reasonable anisotropic
thick shell.
For simplicity, we study the case $G=0$ ($t=t_B$). Hence, the matching
conditions (\ref{9})-(\ref{10}) can be fulfilled by taking, for example
\begin{eqnarray}
& &R={\eta}\;t^{\frac{2}{3}}Y^2+\overline{a}(t)\eta\;{\tilde{Y}}^2,\label{23}\\
& &f(\eta)=Y^2+\sqrt{1+{\eta}_{2}^{2}}{\tilde{Y}}^2,\label{24}\\
& &Y=1-\frac{{(\eta-{\eta}_1)}^2}{{({\eta}_2-{\eta}_1)}^2},\nonumber\\
& &\tilde{Y}=Y=1-\frac{{({\eta}_2-{\eta})}^2}{{({\eta}_2-{\eta}_1)}^2},\nonumber
\end{eqnarray}
where, without loss of generality, we have taken (only for this section!):
\begin{equation} 
a_{Bi}(t_{rec})=1,\;t_{Bi}=t_{rec}=1,\;\eta=\sinh{\eta}_F,\;{\eta}_B=\eta,
\label{atom}
\end{equation}
and ${\eta}_1, {\eta}_2$ denote the location of the comoving shell.
Furthermore, we have set in (\ref{atom}) the scale of times to be the unity at the
recombination era.\\
Regularity conditions impose that $R>0, R_{,\eta}>0$. By performing,
after fixing the time, a Taylor expansion near the boundaries of the thick
shell, we have
\begin{eqnarray}
& &E(t,\eta\simeq{\eta}_1)=\frac{4}{3t^2}+o(1),\label{28}\\
& &E(t,\eta\simeq{\eta}_2)=\frac{3{\overline{a}}_{,t}-1}{{\overline{a}}^2}+o(1)\simeq
\frac{2}{t^2}+o(1),\nonumber\\
& &P_{\eta}(t,\eta\simeq{\eta}_1)=s{({\eta}-{\eta}_1)}^2+o(1),\nonumber\\
& &P_{\eta}(t,\eta\simeq{\eta}_2)=\frac{2}{{\eta}^2}\frac{({\eta}_2-\eta)}
{{\overline{a}}^2}+o(1)\nonumber\\
& &P_{\perp}(t,\eta\simeq{\eta}_1)=s{\eta}_1(\eta-{\eta}_1)+o(1),\nonumber\\
& &P_{\perp}(t,\eta\simeq{\eta}_2)=-\frac{2{\overline{a}}_{,t,t}\overline{a}+
{\overline{a}}_{,t}^{2}}{{\overline{a}}^2}\simeq-\frac{1}{t^2}+o(1)\nonumber\\
& &s=\frac{4t^{-\frac{8}{3}}}{9{\eta}_{1}^{2}{({\eta}_2-{\eta}_1)}^2}[t^{\frac{4}{3}}
(18\sqrt{1+{\eta}_{2}^{2}}-9)+4\overline{a}{\eta}_{1}^{2}-\nonumber\\
& &-18{\eta}_{1}^{2}t^2{\overline{a}}_{,t,t}-12{\eta}_{1}^{2}t{\overline{a}}_{,t}],\nonumber
\end{eqnarray}
where the symbol $\simeq$ on the right hand side
means that the expression is evaluated for
$t>>1$. From the equations (\ref{28}), although the tangential pressure can have a negative value
near $\eta={\eta}_2$, the energy conditions follow near the boundaries of the thick shell.
This means that physically reasonable thick shells can be built with our matching
conditions. It should be stressed again that the junction conditions
(\ref{8})-(\ref{11}) only fix
the behaviour on the boundaries $(1)-(2)$.\\
In the next three sections we apply the technology developed above to physically
interesting situations.

\section{Case with $G=0$}
First of all, we must integrate along the past null cone inward. Generally, we have 
there an equation given by
\begin{equation}
dT=-\frac{A(T,\eta)}{f(\eta)}d\eta,
\label{29}
\end{equation}
where $t_B=\tau=t=T$.\\
Following  C{\ac{e}}l{\ac{e}}rier (see \cite{15,16}), we obtain
for the redshift $z$
\begin{equation}
\frac{dT}{dz}=-\frac{A(T(\eta),\eta)}{(1+z)A_{,T}(T(\eta),\eta)}.
\label{30}
\end{equation}
We are interested in the determination of the distance-redshift formula
for the dimming galaxies. Therefore, from the third of equations
(\ref{17}), we get
\begin{equation}
d_L={(1+z)}^2\overline{a}(t)\sinh{\eta}_F.
\label{31}
\end{equation}
Obviously, we must solve equation (\ref{30}) in the three regions
(hyperbolic Friedmann, thick shell and flat Friedmann) up to an observer
located at $z=0$.\\
For $z\in[z_2,z]$, ($A(T(\eta),\eta)=\overline{a}(T), f(\eta)=1$) we have
\begin{equation}
\frac{1+z}{1+z_2}=\frac{\overline{a}(T_2)}{\overline{a}(T)}.
\label{32}
\end{equation}
For $z\in[z_1,z_2]$, we have ($A(T(\eta),\eta)=R_{,\eta}$)
\begin{equation}
\ln\left(\frac{1+z_2}{1+z_1}\right)=-\int^{T_2}_{T_1}
\frac{R_{,\eta,T}}{R_{,\eta}}dT=\gamma(T_2,T_1).\label{33}
\end{equation}
From (\ref{33}) we get
\begin{equation}
\frac{1+z_2}{1+z_1}=e^{\gamma},\label{34}
\end{equation}
where $\gamma>0$ and physical plausibility requires that
$e^{\gamma}$ is of the order of unity ($\geq 1$).\\ 
For the time flow, we read
\begin{equation}
t=t_B=T=\frac{{\overline{\Omega}}_0}
{2{\overline{H}}_0{(1-{\overline{\Omega}}_0)}^{\frac{3}{2}}}
\left(\sinh\xi-\xi\right).
\label{38}
\end{equation}
Finally, thanks to (\ref{38}),
for $z\leq z_1$ we obtain ($A(T(\eta),\eta)=a_B(T), f(\eta)=1$)
\begin{equation}
1+z_1=\frac{a_B(T_0)}{a_B(T_1)}={\left(\frac
{\sinh{\xi}_0-{\xi}_0}{\sinh{\xi}_1-{\xi}_1}\right)}^{\frac{2}{3}},
\label{35}
\end{equation}
with the subscript "0" denoting the actual time at $z=0$. 
By multiplying equations (\ref{32})-(\ref{35}), we get
\begin{equation}
1+z=e^{\gamma}\;\frac{\overline{a}(T_2)}{\overline{a}(T)}\frac{a_B(T_0)}{a_B(T_1)}.
\label{36}
\end{equation}
For ${\eta}_F$ along the past null cone we have
\begin{equation}
{\eta}_F={\xi}_0-\xi\;\;,\;\;\xi\leq {\xi}_0.
\label{37}
\end{equation}
As a result, since ${\overline{\Omega}}(\xi)=\frac{2}{1+\cosh\xi}$, 
equation (\ref{36}) becomes
\begin{equation}
1+z=2F(1+z_1)\frac{(1-{\overline{\Omega}}_1)}{
{\overline{\Omega}}_1(\cosh\xi-1)},
\label{39}
\end{equation}
where "$1$" refers to the time $T_1$ (or ${\xi}_1$) and
\begin{equation}
F=e^{\gamma}\frac{(\cosh{\xi}_2-1)}{(\cosh{\xi}_1-1)}\label{40}.
\end{equation}
Physical plausibility requires that $F$ be of order of unity.
Thanks to (\ref{3}) and (\ref{39}), 
the formula (\ref{31}) becomes
\begin{equation}
d_L=\frac{{\overline{\Omega}}_0(1+z)(1+z_1)}
{{\overline{H}}_0{(1-{\overline{\Omega}}_0)}^{\frac{3}{2}}}
\frac{F(1-{\overline{\Omega}}_1)}{
{\overline{\Omega}}_1}\sinh({\xi}_0-\xi).
\label{41}
\end{equation}
Equations (\ref{39}) and (\ref{41}) hold for
$z\geq z_2$ and obviously the junction conditions 
only imply the continuity of $d_L$ and not its analyticity when crossing
the boundaries of the model.
This obviously applies to any inhomogeneous model built by
matching two or more metrics (as for example in \cite{10,12,20BB}).
Nevertheless, to make contact
with the astrophysical data at intermediate redshifts,
a central observer can formally expand 
expression (\ref{41}) in a Taylor series in the range
$z_1<<1, z_2-z_1<<1$ (${\overline{\Omega}}_0\simeq{\overline{\Omega}}_1,
F\simeq 1$).
To the first order in $z$ we formally obtain
\begin{eqnarray}
& &d_L=\frac{z}{H_{obs}}+o(z),\label{42}\\
& &H_{obs}=\frac{\epsilon\;{\overline{H}}_0{(1-{\overline{\Omega}}_0)}^{\frac{3}{2}}
{\overline{\Omega}}_1}
{F(1-{\overline{\Omega}}_1)(1+z_1){\overline{\Omega}}_0\;P},\nonumber\\
& &P=\sqrt{F\epsilon(1-{\overline{\Omega}}_1)(1+z_1)},\nonumber\\
& &\epsilon=-{\overline{\Omega}}_1[F(1+z_1)-1]+F(1+z_1).\nonumber
\end{eqnarray}
Therefore,  in our inhomogeneous universe, after writing
the correct matching conditions, a central observer extrapolates an effective 
Hubble flow given by (\ref{42}). It is worth noticing that,
in the limit 
$F=1, {\overline{\Omega}}_1\rightarrow{\overline{\Omega}}_0, z_1=0,
\epsilon=1$, in which the full space-time is composed only with the hyperbolic
Friedmann metric,
we have $H_{obs}\rightarrow{\overline{H}}_0$,
a correct result. 
Furthermore, the inequality $H_{obs}\neq{\overline{H}}_0$ in a 
general inhomogeneous universe is compatible with the fact that in such space-times
we have not a unique definition of an observed Hubble flow
(see \cite{27}): a direct way is to 
infer its value by a formal Taylor expansion (if this is possible)
near the observer.
The extrapolated central deceleration parameter $q_0$ is given by
\begin{equation}
q_0=-H_{obs}\frac{d^2}{dz^2}\left(d_L(z=0)\right)+1.
\label{43}
\end{equation}
Note that, from equation (\ref{35}), we could express ${\overline{\Omega}}_1$ as 
a function of $z_1, {\xi}_0$, although this is not necessary for our purposes.
Equation (\ref{43}), thanks to (\ref{41}), gives
\begin{equation}
q_0=\frac{{\overline{\Omega}}_1}{2\epsilon}.
\label{44}
\end{equation}
From equation (\ref{42}) we must have $\epsilon>0$ ($H_{obs}>0$)
and therefore $q_0\geq 0$ , for all times $T$ and no "formal" acceleration is perceived by the central
observer by considering the distance-redshift function.
At early times ($\xi\simeq 0$) we have $q\rightarrow\frac{1}{2}$ and 
$q\rightarrow 0^{+}$ asymptotically (for ${\xi}_1\rightarrow\infty$). 

\section{Uniform Hubble flow}
The case of a uniform Hubble flow has been studied in \cite{10,12} in the context
of the Buchert equations with backreaction (\cite{5}). In \cite{10,12} the overdensity evolves
asymptotically as an Einstein-de Sitter space-time, while the underdensity as 
a Milne universe. As a result, our model can be considered as representing the 
far future limit ($t\rightarrow\infty$) of \cite{10,12}, where backreaction is asymptotically
vanishing. 
With the uniform Hubble gauge we have
\begin{equation}
\overline{H}=H_B,\;\alpha=1,\;J(\xi)=
\frac{dt}{dt_B}=\frac{3}{2}\frac{(1+\cosh\xi)}{(2+\cosh\xi)}.
\label{45}
\end{equation}
The calculations are similar to the ones of the last section. 
However, the central observer measures the redshift with respect to its proper time
$t_B$. As a result, along the past null cone we have
\begin{eqnarray}
& &dt_B=-\frac{\overline{a}}{J(\xi)}d{\eta}_F\rightarrow{\eta}_F={\xi}_0-\xi\nonumber\\
& &\frac{1+z}{1+z_2}=\frac{J(\xi)}{J({\xi}_2)}
\frac{\overline{a}({\xi}_2)}{\overline{a}(\xi)}.\label{47}
\end{eqnarray}
Instead of the equation (\ref{33}) we read
\begin{eqnarray}
& &\frac{1+z_2}{1+z_1}=e^{\gamma}\simeq\frac{J({\xi}_2)}{J({\xi}_1)}
\frac{(1+{\overline{z}}_2)}{(1+{\overline{z}}_1)},\label{48}\\
& &\frac{(1+{\overline{z}}_2)}{(1+{\overline{z}}_1)}=
-\int^{{\tau}_2}_{{\tau}_1}\frac{e^{\frac{G}{2}}}{R_{,\eta}}
{\left(\frac{R_{,\eta}}{e^{\frac{G}{2}}}\right)}_{,\tau}d\tau,\label{49}
\end{eqnarray}
${\overline{z}}_1, {\overline{z}}_2$ being the redshifts measured by a comoving
observer with time $\tau$. The approximation ($\simeq$) in (\ref{48}) has been 
given as an example and is valid when ${\xi}_2\simeq{\xi}_1$. It does not enter
in the calculations of this section. The expression (\ref{31}) becomes
\begin{equation}
d_L={(1+z)}^2\frac{\overline{a}(t)}{J(\xi)}\sinh({\xi}_0-\xi).
\label{50}
\end{equation}
Concerning the relation between $t_B$ and $t$ we get
\begin{equation}
t_B=\frac{{\overline{\Omega}}_0}
{3{\overline{H}}_0{(1-{\overline{\Omega}}_0)}^{\frac{3}{2}}}
\frac{{(\cosh{\xi}-1)}^{\frac{3}{2}}}{{(1+\cosh{\xi})}^{\frac{1}{2}}}.
\label{51}
\end{equation}
Instead of equation (\ref{35}) we have
\begin{equation}
1+z_1={\left(\frac{t_{B0}}{t_{B1}}\right)}^{\frac{2}{3}}.
\label{35b}
\end{equation}
After defining 
\begin{equation}
e^{\gamma}=F\frac{(\cosh{\xi}_1-1)}{(\cosh{\xi}_2-1)}
\frac{(2+\cosh{\xi}_1)}{(2+\cosh{\xi}_2)}
\frac{(1+\cosh{\xi}_2)}{(1+\cosh{\xi}_1)},
\label{52}
\end{equation}
and with the same technique of the last section, we obtain
\begin{eqnarray}
& &\cosh\xi=-\frac{1}{2}+\frac{Q}{2(1+z)}+\nonumber\\
& &+\frac{\sqrt{9{(1+z)}^2+2Q(1+z)+Q^2}}{2(1+z)},\label{53}\\
& &Q=\frac{F}{{\overline{\Omega}}_1}(1+z_1)(1-{\overline{\Omega}}_1)
(2+{\overline{\Omega}}_1),\label{54}\\
& &d_L=\frac{(2+\cosh\xi)(\cosh\xi-1)\sinh({\xi}_0-\xi)}
{3{\overline{H}}_0{(1-{\overline{\Omega}}_0)}^{\frac{3}{2}}
(1+\cosh\xi)}{(1+z)}^2.\label{55}
\end{eqnarray}
Performing a formal Taylor expansion of (\ref{55}) at $z=0$, we again get an
effective observed $H_{obs}$, and by means of equation (\ref{43}) we can
obtain the central deceleration parameter. In any case, we always have
$q=\frac{1}{2}$ at early times. Furthermore, the extrapolated parameter $q_0$
remains positive and approaches zero 
as follows: $q({\xi}_1\rightarrow\infty)\rightarrow\frac{7}{Q^2}$.
As a result, in presence of a uniform Hubble flow 
$q_0$ goes to zero more rapidly than in the case $G=0$
(see equation (\ref{44})), albeit always
from positive values.
In practice, we recover the results of 
\cite{11}, but imposing the correct matching conditions.

\section{The general case}
In the general case $\overline{H}=\alpha\;H_B$. Hence, the relation
between $t_B$ and $t$ becomes
\begin{equation}
t_B=\alpha(\xi)\frac{{\overline{\Omega}}_0}
{3{\overline{H}}_0{(1-{\overline{\Omega}}_0)}^{\frac{3}{2}}}
\frac{{(\cosh{\xi}-1)}^{\frac{3}{2}}}{{(1+\cosh{\xi})}^{\frac{1}{2}}}.
\label{56}
\end{equation}
For the lapse factor $J(\xi)$, we obtain
\begin{eqnarray}
& &J(\xi)=\frac{dt}{dt_B}=\frac{3}{2}\beta(\xi),\label{57}\\
& &{\alpha}_{,\xi}\frac{(\cosh\xi-1)}{\sinh\xi}+\alpha
\frac{(2+\cosh\xi)}{(1+\cosh\xi)}={\beta}^{-1}(\xi).\nonumber
\end{eqnarray}
To explore the case with a non-uniform Hubble flow with $G\neq 0$ we
can take
\begin{equation}
\beta(\xi)=\frac{(1+\cosh\xi)}{(A\cosh\xi+3-A)},\;\;\;A>0.
\label{58}
\end{equation}
The case with $A=1$ has been studied in the section above. The
calculations are similar to the ones of the section 6 and after posing
\begin{equation}
e^{\gamma}=\frac{1+z_2}{1+z_1}=\frac{F{\overline{\Omega}}_2(1-{\overline{\Omega}}_1)}
{{\overline{\Omega}}_1(1-{\overline{\Omega}}_2)}
\frac{(2A-2A{\overline{\Omega}}_1+3{\overline{\Omega}}_1)}
{(2A-2A{\overline{\Omega}}_2+3{\overline{\Omega}}_2)},
\label{59}
\end{equation}
we get
\begin{eqnarray}
& &d_L=\frac{B(1+z){\overline{\Omega}}_0}{3{\overline{H}}_0
{(1-{\overline{\Omega}}_0)}^{\frac{3}{2}}}\sinh({\xi}_0-\xi),\label{60}\\
& &1+z=\frac{B(1+\cosh\xi)}
{(A\cosh\xi+3-A)(\cosh\xi-1)},\label{61}\\
& &B=\frac{F(1+z_1)(1-{\overline{\Omega}}_1)
(2A-2A{\overline{\Omega}}_1+3{\overline{\Omega}}_1)}
{{\overline{\Omega}}_1}.\label{62}
\end{eqnarray}
The expressions
for $H_{obs}$ and $q_0$ are rather cumbersome. However, some general remarks
can be done on the behaviour of the extrapolated central deceleration parameter
$q_0$ at different values of the parameter $A$.
First of all, $\forall A\in(0,\infty)$,
$q_0(\xi=0)=\frac{1}{2}$. Furthermore,
$\forall A\in[1,\infty)$, the 
parameter $q_0$ is always positive and asymptotically 
$q_0(\xi\rightarrow\infty)\rightarrow 0^{+}$. Conversely,
$\forall A\in(0,1)$, the parameter $q_0$ becomes negative at some value of $B$,
and after an absolute minimum, reaches $0^{-}$ asymptotically but from
negative values. As an example, for $A=\frac{1}{2}$, $q_0=0$ for
$B_0\simeq 5.3$ and for $B\simeq 10$ we have $q_{0min}\simeq -0.075$.
For $A=\frac{1}{3}$, $q_0=0$ for $B_0\simeq 4.2$ with 
$q_{0min}\simeq -0.16$ at $B\simeq 8$. For $A=\frac{1}{5}$,
$q_0=0$ for $B_0\simeq 3.6$ with $q_{0min}\simeq -0.31$ at $B\simeq 6$.
For $A=\frac{1}{9}$, $q_0=0$ for $B_0\simeq 3.2$ and
$q_{0min}\simeq -0.51$ at $B\simeq 5$. In any case, 
$\forall A\in(0,1)$ it is always possible to have, from the
equation (\ref{62}), reasonable values for $F$ ($\simeq 1$, with
$z_2-z_1<<1$) and ${\overline{\Omega}}_0$ 
compatible
with an apparent acceleration at some later time calculated by means of the
equation (\ref{35}), provided that $B>B_0$. The inequality
$B>B_0$ imposes a constraint on 
${\overline{\Omega}}_1\simeq {\overline{\Omega}}_0$. As an example,
for $F=1.1, z_1=0.01$ and $A=\frac{1}{2}$ we have an accelerated universe
if and only if ${\overline{\Omega}}_1\simeq {\overline{\Omega}}_0<0.24$, 
while for $A=\frac{1}{2}$ and $F(1+z_1)=1, z_1<<1$, we have
${\overline{\Omega}}_1\simeq {\overline{\Omega}}_0<0.22$.
Further, for $A=\frac{1}{9}, F=1.1, z_1=0.01$
we have formal acceleration when 
${\overline{\Omega}}_1\simeq {\overline{\Omega}}_0<0.25$.
It is also possible to mimic for $A\leq\frac{1}{9}$ a value
for $q_0$ compatible with the actual observations.
Furthermore, note that the models with $A\in(0,1)$
represent the case with $\alpha(\xi)<1$., i.e.
$H_B>\overline{H}$(this can be see by noting that the equation 
(\ref{57}) has, in the limit $\xi>>1$, the tracker solution 
$\alpha=A$). As a result, the clock effects depicted in this paper can
mimic a model with a large underdensity surrounded by an overdensity
(see \cite{20BB}). 
Concluding, if the clock effects depicted in this paper (and in 
\cite{10,12}) there exist in the  real universe, the actual data at intermediate redshifts
are in agreement with a non-uniform Hubble flow.

\section{Conclusions}
We built a model for the universe without 
dark energy, by means of an exact spherically symmetric solution
taking into account the observed inhomogeneous universe.
The main purpose of this paper is to show how the non-uniform time flow
depicted in \cite{10,12} can be obtained within an exact solution 
of Einstein's equations by imposing the correct matching conditions
required by general relativity. In this sense, since the backreaction is
absent in our model,
our approach is different
from Wiltshire's, where the backreaction is analyzed in a statistical
("Copernican") model within the Buchert formalism.\\
The model is composed of three regions, a central flat Friedmann metric, 
an hyperbolic Friedmann zone and eventually a bulk LTB hyperbolic
metric, according to WMAP. Within our exact solution , it is shown that,
after a "formal" Taylor expansion of the distance-redshift relation near 
the observer and by imposing the correct
matching conditions, a uniform Hubble "gauge" (present in \cite{10,12})
does not lead to an "apparent" acceleration as extrapolated from
the redshift-distance relation. 
In a purely spherically symmetric universe, such an acceleration can only be obtained
with a non-uniform Hubble flow.
Furthermore, in our model we have a parameter ($A$ in the paper)
at our disposal that permits us to obtain a large amount of "apparent" acceleration which is consistent
with the other parameters of the model (for example the thickness $z_2-z_1<<1$).
Furthermore, in the presentation given in \cite{10,12} the
physical mechanism that can generate a non-uniform time flow  is not yet clear. 
In fact it is always possible
to build an inhomogeneous universe with a global cosmic time by means, for example, of 
LTB metrics. An exact formulation requires that the junction conditions are
fulfilled only by means of a thick shell. We have shown a possible
link between a non-uniform time flow and a radial energy flux present in the energy-momentum
tensor of the thick shell. Hence, this "heat flow" term can give a possible physical
explanation for the clock effects depicted in \cite{10,12} (if they exist!). 
The introduction of thick shells can be useful to explore exact models
obtained by glueing different Friedmann metrics. An anisotropic thick shell
is certainly an unusual choice, 
but this  alleviates the drawbacks of a whole
universe filled with an exotic dark energy.
It should be noted that the radial flux energy vanishes on the boundaries of
the thick shell. In fact the radial symmetry inhibits a "heat flow"
between the flat central Friedmann metric and the "dimming" hyperbolic
Friedmann one. A more realistic model could be obtained by relaxing the 
spherical symmetry and susbstituing the Friedmann metrics with more general ones,
admitting a non-vanishing energy flux. Unfortunately, nowadays
such metrics are not at our disposal.
In any case, our calculations can suggest an improvement over Wiltshire's model.
In fact, Wiltshire's paper neglects the shear, but this encodes 
fundamental informations related to the variation of non-local gravitational energy,
which is a fundamental ingredient in \cite{10,12}.
As is well know, it is not a simple task to relate the shear to
physical observable quantities (see \cite{31bis}). To this purpose, Wiltshire's
model could be amended by taking
\begin{eqnarray}
& &{(<\theta>)}_{fi}=0,\;{(<{\sigma}^2>)}_{fi}=0,\label{ret}\\
& &{(<\theta>)}_{s}=3H_{s},\;{(<{\sigma}^2>)}_{s}\neq 0,\\
& &{(<\theta>)}_{v}>0,\;{(<{\sigma}^2>)}_{v}=0,
\end{eqnarray}
where, following the notation of \cite{10}, "{\it fi}" stands for "finite-infinity" 
and "{\it v}"
for "voids", $H_s$ is the averaged Hubble parameter for the thick shell. Hence, as
suggested by the matching conditions, a third scale between "{\it fi}" 
and "{\it v}" with
a non-vanishing shear is introduced: this is the scale at which variations in the
flux energy are appreciable.\\
Finally, note that the dark energy appears in two "phase transitions"
for the universe: the formation of the big structures and the end of it.
Hence, since cosmic strings and superstrings in the context of the
M-theory
are supposed to 
have acted during the inflation epoch to give (in principle) observable
effects a later times, 
a link between them and the thick structures
depicted in this paper  can be suggested (see \cite{cosmic}). 

\end{document}